\documentstyle[epsf]{article}

\newcommand{\re}{\par\hangindent=0.5cm\hangafter=1\noindent}

\newcommand{\bm}{\boldmath}
\newcommand{\bmv}{\mbox{\bm $v$}}

\newcommand{\bmnabla}{\mbox{\bm $\nabla$}}
\newcommand{\bmF}{\mbox{\bm $F$}}

\begin{document}
%

\baselineskip 14pt

\begin{center}
{\large {\bf Effect of Radiation Drag on Hoyle-Lyttleton Accretion}} \\
  
\rm  Tomomi {\sc Nio},$^{1*}$ Takuya {\sc Matsuda},$^1$ 
and Jun {\sc Fukue}$^2$ \\
$^1$ {\it Department of Earth and Planetary Sciences,
          Kobe University, Kobe 657-8501}\\
$^2$ {\it  Astronomical Institute, Osaka Kyoiku University,
           Asahigaoka, Kashiwara, Osaka 582-8582}\\
{\it E-mail(TN): nio@stp.isas.ac.jp}\\
\end{center}

$^*$
Present address:
The Institute of Space and Astronautical Science, 
Yoshinodai, Sagamihara, Kanagawa 229-8510
and
Department of Earth and Planetary Physics,
University of Tokyo, 
Hongo, Bunkyo-ku, Tokyo 113-8654

%


\begin{center}
{\bf Abstract}
\end{center}

Hoyle-Lyttleton type accretion is investigated, 
by taking account of 
not only the effect of radiation pressure 
but the effect of radiation drag.
We calculate the trajectories of particles for three cases:
only the effect of gravity is considered (case A);
the effect of radiation pressure is taken into account (case B);
the effect of radiation drag as well as radiation pressure
is taken into account (case C).
The accretion radii for former two cases are 
$2GM/v_{\infty}^2$ for case A
and $2GM(1-\Gamma)/v_{\infty}^2$ for case B,
where $M$ is the mass of the accreted object,
$v_{\infty}$ the relative velocity, and
$\Gamma$ the normalized luminosity of the accreted object.
We found that the accretion radius for case C is 
in between those of cases A and B 
under the present approximation;
i.e.,
the accretion radius decreases due to radiation pressure
while it increases due to radiation drag.
In addition, the accretion radius for case C
becomes larger as the incident velocity becomes fast.
The effect of radiation drag becomes more and more important
when the velocity of the incident particle is 
comparable to the light speed.
%




\begin{flushleft}
{\bf 1. Introduction}
\end{flushleft}

Accretion is a phenomenon, in which gas is attracted by 
a gravitating astrophysical object and accretes on it.
In this process potential energy is liberated, and
the accreting gas becomes hot and emits radiation.
If the accreted body is a compact object 
such as a neutron star or a black hole,
significant fraction of the rest mass energy of
the accreting gas is liberated,
and the efficiency of energy conversion exceeds
that of nuclear energy. 
Thus, the accretion process can be 
a major source of energy in astrophysics.

In the simplified model of the axisymmetric accretion 
introduced by Hoyle and Lyttleton (1939), 
the field of a gravitating body acts on 
surrounding streams of particles,
which has a uniform flow upsteam at infinity.
These particles move on hyperbolic orbits,
which intersect downstream of the point mass 
on the axis of symmetry (the accretion axis). 
Pressure forces are everewhere neglected, 
although the intersecting particles are allowed to collide 
and coalesce on the accretion axis. 
Particles colliding close enough to the point mass lose 
enough of their kinetic energy to be trapped by the central object. 

Bondi and Hoyle (1944) replaced the infinitely-thin accretion axis 
by a broader accretion column, in which pressure forces play a role
(see also Davis, Pringe 1980).
Bondi (1952) also considered a spherical accretion
without bulk motion of a gravitating object,
taking account of the gas pressure
(see Treves et al. 1988 for a review).
In the Hoyle-Lyttleton type accretion
the effect of gas pressure is taken into account numerically 
by several researchers
(Hunt 1971; Shima et al. 1985; Fryxell et al. 1987;
Ho et al. 1989; Koide et al. 1991).
In these calculations, however, 
the effect of radiation pressure was neglected.
The effect of radiation pressure was taken into account by, e.g.,
Taam et al. (1991).
Due to radiation pressure,
the accretion radius and the accretion rate
are significantly reduced from the Hoyle-Lyttleton estimate.

In these studies
they have neglected the effect of {\it radiation drag}
(such as Compton drag).
In the ionized gas the radiation drag force becomes important
if the velocity of gas is 
comparable to the light speed
and the luminosity of the central object is 
comparable to the Eddington luminosity.
Since such a condition was supposed to be rarely fulfilled 
in realistic astrophysical situations,
the effect of radiation drag has been neglected so far.
Radiation drag is also important,
if the specific cross-section --- ``opacity'' ---
of particles is sufficiently large like, e.g., dust.

Of course, since the discovery of Poynting-Robertson effect
(Poynting 1903; Robertson 1937),
the effect of radiation drag on particles moving around
the radiating object has been studied by several authors.
For example, Guess (1962) examined the Poynting-Robertson effect
for the case of a finite spherical source.
He also examined the particle trajectories in details.
Carroll (1990) also examined the Poynting-Robertson effect
more generally.
In the field of accretion disks, furthermore,
the effect of radiation drag was extensively investigated recently,
since in the inner region of accretion disks
the radiation field is so intense and the rotation velocity
is so high that radiation drag becomes important
(see, e.g., 
Fortner et al. 1989;
Meyer, Meyer-Hofmeister 1994;
Fukue, Umemura 1995;
Tajima, Fukue 1996; Watanabe, Fukue 1996;
see also Kato et al. 1996).

%
In Hoyle-Lyttleton type accretion, 
the effect of radiation drag has not been considered yet.
However, there can be a situation
in which a luminous source moves at a high velocity
and radiation drag becomes important.
For example,
in the case of supernova explosions
a neutron star can be ejected from a binary system
at a high speed comparable to the light speed.
In compact binaries
a neutron star or a black hole surrounded by accretion disks,
which rotates at a high orbital velocity,
can feed gas from companion winds.
In astrophysical jets, 
whose speed is of the order of the light speed,
a gravitating object can be located in the jet flow.
Moreover, since the effect of radiation drag is 
a fundmental process in astrophysics,
it is worthwhile to investigate it.

In the next section 
we describe an axisymmetric accretion,
the radiation fields, and the basic equations.
Numerical results are presented in section 3.
Final section is devoted to the concluding remarks.

\begin{flushleft}
{\bf 2. Model}
\end{flushleft}

\begin{flushleft}
{\it 2.1. Hoyle-Lyttleton Axisymmetric Accretion}
\end{flushleft}

As we discussed in the introduction, 
the axisymmetric accretion is modelized as follow.
Let us suppose that
a graviating body with mass $M$ and luminosity $L$ 
is placed in a flow, 
of which velocity at the infinite upstream is 
$v_{\infty}$ and of which density is $\rho_{\infty}$.
If the velocity $v_{\infty}$ is much larger 
than the sound speed $c_{\infty}$ of the gas at infinity
(or if the Mach number of the upstream flow is much larger than unity),
then the pressure effect can be neglected
except for the very center.
If we compare the kinetic energy of the gas, $v_{\infty}^2/2$,
with the potential energy of the gas, $GM/r$,
we obtain the {\it Hoyle-Lyttleton accretion radius} $R_{\rm HL}$:
\begin{equation}
        R_{\rm HL}=\frac{2GM}{v_{\infty}^2}
\end{equation}
(More precisely,
the accretion radius is derived from the escape condition
on the accretion axis, as described later).
Inside the accretion radius
the potential energy exceeds the kinetic energy
and the gas is trapped by the gravitating object.
By a simple argument, Hoyle and Lyttleton (1939) showed that 
a gas with the impact parameter less than $R_{\rm HL}$ 
accretes onto the gravitating body.
Thus, the rate of mass accretion, ${\dot M}_{\rm HL}$,
on to the gravitating object is
\begin{equation}
    {\dot M}_{\rm HL} = \pi R_{\rm HL}^2 \rho_{\infty} v_{\infty}
           = \frac{4\pi\rho_{\infty}G^2 M^2}{v_{\infty}^3}.
\end{equation}
If we know the accretion rate, 
we can estimate the (accretion) luminosity of the radiation 
emitted from the compact object. 
Therefore, the knowledge on the accretion radius is important 
to compare theories with observations.

Bondi (1952) considered a spherical accretion 
and obtained the so-called Bondi radius $R_{\rm B}$:
\begin{equation}
   R_{\rm B}=\frac{GM}{c_{\infty}^2},
\end{equation}
where $c_{\infty}$ is the sound speed of gas at infinity.
Hence, if we take account of the gas pressure,
the accretion radius and accretion rate are modified respectively as
\begin{equation}
   R_{\rm HLB}=\frac{2GM}{v_{\infty}^2+c_{\infty}^2},
\end{equation}
\begin{equation}
    {\dot M}_{\rm HLB} 
           = \frac{4\pi\rho_{\infty}G^2 M^2}
            {(v_{\infty}^2+c_{\infty}^2)^{3/2}},
\end{equation}
considering the results of numerical simulations
(Bondi 1952; Shima et al. 1985).

Later, Taam et al. (1991) took into account 
the effect of the radiation pressure force,
and obtained the accretion radius modified by the radiation pressure:
\begin{equation}
   R_{\rm HL}^{\rm R}=\frac{2GM(1-\Gamma)}{v_{\infty}^2},
\label{eqn:bondi}
\end{equation}
when the gas is fully ionized and transparent.
Here, $\Gamma$ is the indication of luminosity strength, 
i.e., the ratio of the luminosity $L$ of the gravitating object 
to the Eddington luminosity 
$L_{\rm E}$ ($=4 \pi cGMm_{\rm p}/\sigma_{\rm T}$):
\begin{equation}
       \Gamma = \frac{L}{L_{\rm E}},
\end{equation}
where $m_{\rm p}$ is the proton mass
and $\sigma_{\rm T}$ the Thomson scattering cross-section.
The reason, why the effect of radiation pressure is expressed
by such a simple formula as equation (\ref{eqn:bondi}),
is because the radiation pressure decreases as $r^{-2}$,
which is the same dependence as that of gravity.
The formula (\ref{eqn:bondi}) means the following facts:
(i) If the luminosity of the gravitating object
is sub-Eddington, the accretion radius reduces
by a factor $(1-\Gamma)$
compared with the Hoyle-Lyttleton radius.
(ii) If the luminosity is super-Eddington,
the accretion cannot take place and
the gas particles will be scattered
by the radiation pressure
(the radiative Rutherford scattering).

\begin{flushleft}
{\it 2.2. Radiation Fields of the Central Source}
\end{flushleft}

Let us assume that
the luminous gravitating object is located at the origin,
and it is immersed in the uniform gas flow.
The total luminosity of the central object is $L$.
We also assume that the gas is fully ionized and transparent 
to the radiation from the object.
We adopt the cylindrical coordinates $(r, \varphi, z)$,
where the $z$-axis is in
the direction of the gas particle at the upstream.

At a point whose radial distance from the central object is
$R$ ($= \sqrt{r^2+z^2})$,
the radiation energy density $E$,
the radiative flux $F^{R}$ in the $R$-direction,
and the diagnal component of the radiation stress tensor $P^{RR}$ 
are expressed respectively as
\begin{eqnarray}
   E      &=& \frac{F^R}{c} = \frac{1}{c}\frac{L}{4\pi R^2}, \\
   F^R    &=& \frac{L}{4\pi R^2}, \\
   P^{RR} &=& E = \frac{1}{c}\frac{L}{4\pi R^2}, 
\end{eqnarray}
in the region where the central object is treated as a point source
(other components vanish).
It should be noted that
if the effect of the finite source size is taken into account,
the radiation energy density is enhanced 
in the very vicinity of the central object
(Guess 1962; Carroll 1990).

\begin{flushleft}
{\it 2.3. Basic Equations}
\end{flushleft}

In order to make the problem tractable,
let us assume that the flow is axisymmetric 
about the axis penetrating the center of the gravitating body 
and parallel to the incident flow (accretion axis).
We neglect the gas pressure, the viscosity,
the heating, the cooling, and the magnetic field.

The motion of the gas particle in the radiation fields
is described by
\begin{equation}
\frac{d\bmv}{dt}
     =-\bmnabla \phi + \frac{\sigma_{\rm T}}{mc}
       (\bmF-E \bmv -{\bf P} \otimes \bmv),
\label{vector}
\end{equation}
upto the first order of $v/c$,
where $\bmv$ is the particle velocity,
$\phi$ the gravitational potential,
$m$ the particle mass (proton mass for the normal plasma),
$E$ the radiation energy density,
$\bmF$ the radiative flux vector,
and {\bf P} the radiation stress tensor
(Hsieh, Spiegel 1976; Fukue et al. 1985;
see also Kato et al. 1998).
The terms
$(-E\bmv-{\bf P}\otimes \bmv)$
express the radiation drag force,
which is our main concern in the present paper.

In the cylindlical coordinates $(r, \varphi, z)$,
the equation (\ref{vector}) is expressed as
{
\setcounter{enumi}{\value{equation}}
\addtocounter{enumi}{1}
\setcounter{equation}{0}
\renewcommand{\theequation}{\theenumi\alph{equation}}
\begin{eqnarray}
    \frac{\displaystyle dv_r}{\displaystyle dt} 
    &=& -\frac{\displaystyle GMr}{\displaystyle R^3}
        +\frac{\displaystyle \sigma_{\rm T}}{\displaystyle mc}
         \left[F^R \frac{\displaystyle r}{\displaystyle R}
               - Ev_r
               - P^{RR}\frac{\displaystyle r}{\displaystyle R}
         \left(v_r \frac{\displaystyle r}{\displaystyle R} 
             + v_z \frac{\displaystyle z}{\displaystyle R} \right)
         \right], 
         \\
    \frac{\displaystyle dv_z}{\displaystyle dt} 
    &=& -\frac{\displaystyle GMz}{\displaystyle R^3}
        +\frac{\displaystyle \sigma_{\rm T}}{\displaystyle mc}
         \left[F^R \frac{\displaystyle z}{\displaystyle R}
               - Ev_z
               - P^{RR}\frac{\displaystyle z}{\displaystyle R}
         \left(v_r \frac{\displaystyle r}{\displaystyle R}
             + v_z \frac{\displaystyle z}{\displaystyle R} \right)
         \right],  
\label{eqn:basiceq0}
\end{eqnarray} 
retaining the non-zero components of radiation fields.
Let us normalize the length and the time by
$r_{\rm g}$ and $r_{\rm g}/c$, respectively,
where $r_{\rm g}$ is the Schwarzschild radius
of the central object.
Subtituting the expressions for $E$, $F^R$, and $P^{RR}$,
and normalizing the resultant equations,
we finally have the following equations:
\setcounter{equation}{\value{enumi}}
\setcounter{enumi}{\value{equation}}
\addtocounter{enumi}{1}
\setcounter{equation}{0}
\renewcommand{\theequation}{\theenumi\alph{equation}}
\begin{eqnarray}
    \frac{\displaystyle dv_r}{\displaystyle dt} 
    &=& -\frac{\displaystyle (1-\Gamma)r}{\displaystyle 2R^3}
        -\frac{\displaystyle \Gamma}{\displaystyle 2R^2}
         \left[\left(1+\frac{\displaystyle r^2}{\displaystyle R^2}
               \right)v_r+ 
         \frac{\displaystyle rz}{\displaystyle R^2}v_z \right], 
         \\
    \frac{\displaystyle dv_z}{\displaystyle dt} 
    &=& -\frac{\displaystyle (1-\Gamma)z}{\displaystyle 2R^3}
        -\frac{\displaystyle \Gamma}{\displaystyle 2R^2}
         \left[\frac{\displaystyle rz}{\displaystyle R^2}v_r+
         \left(1+\frac{\displaystyle z^2}{\displaystyle R^2}\right)v_z
         \right].  
\label{eqn:basiceq}
\end{eqnarray}
Here, the symbols for the dimensionless variables,
say ``$\hat{v}$'', are dropped. 
\setcounter{equation}{\value{enumi}}
}

There are two physical parameters defining the problem: 
$v_{\infty}$ and $\Gamma$.
We consider the following three cases:
\begin{itemize}
\item
Case A:
If we set $\Gamma$ to be zero, we obtain the equations 
of ballistic orbits of a particle mass.
This is just the original Hoyle-Lyttleton accretion.
\item
Case B:
If we neglect the second terms of the right-hand sides of 
equations (13), 
then we obtain the equations of motion of a particle in which 
only the gravity and the radiation pressure are taken into account.
This is the case examined by, e.g., Taam et al. (1991).
\item
Case C:
If we take into account all the terms in equations (13),
we obtain the equations of motion of a particle which is affected
not only by the gravity and the radiation pressure
but also radiation drag.
\end{itemize}

\begin{flushleft}
{\bf 3. Results}
\end{flushleft}

\begin{flushleft}
{\it 3.1. Particle Trajectory}
\end{flushleft}

We first calculate trajectories of particles
by integrating equation (13) 
using the forth-order Runge-Kutta-Gill method.
In cases A and B, as is well known,
particle trajectories are hyperbolae.
The equation of hyperbolic orbits in the polar coordinates is
$r = \ell /(1 + \varepsilon \cos \theta)$,
where
$\ell = b^2 v_\infty^2/[GM(1-\Gamma)]$ and
$\varepsilon = \sqrt{1+b^2 v_\infty^4/[G^2 M^2(1-\Gamma)^2]}$,
$b$ being the impact parameter.
In order to compare the present case C with cases A and B,
and to check the numerical accuracy,
we show the numerical results of all cases.

Figure 1 show the trajetories of gas particles
in the case of $v_{\infty}=0.1c$ for several $\Gamma$;
$\Gamma =0$ in figure 1a
(i.e., the central object does not emit radiation),
$\Gamma=0.5$ in figure 1b,
$\Gamma=1.0$ in figure 1c
(i.e., the luminosity of the central object is the Eddington one),
and $\Gamma=1.2$ in figure 1d.
Short-dashed curves are for case A (gravity only),
chain-dotted ones for case B (gravity and radiation pressure),
and solid ones for case C (including radiation drag).

In figure 1b (sub-Eddington luminosity)
we see the example where
a particle accretes to the central object
if radiation drag is included,
otherwise a particle can escape.

In figure 1c (Eddington luminosity)
the trajectories of particles in cases B and C 
are almost straight lines,
since the effect of gravity is canceled by the radiation pressure.
Precisely speaking,
in case B the trajectory is exactly straight,
while it slightly deviates from the straight line in case C.
This is understood as follows.
In case C including radiation drag,
a particle initially moves in the $z$-direction and
does not move in the $r$-direction
so that there is no radiation drag force
originating from the radiation energy density $E$.
However, there exists the radiation drag force
arising from the radiation stress tensor $P^{RR}$
[see equation (13)],
since $P^{RR}$ interacts with $v_z$.
As a result, there appears a residual (positive) force
in the $r$-direction for $v_z<0$.
This is the case of the trajectory in case C.

In figure 1d (super-Eddington luminosity) 
the particles in cases B and C are expelled 
due to the radiation pressure
(this is just the {\it radiative Rutherford scattering}).
In case C with radiation drag,
the ``scattering'' is apparently strong
compared with that of case B.
This is understood as follows.
In case C
the particle motion in the $z$-direction
is suppressed by the effect of radiation drag,
and the vertical velocity $v_z$ becomes slow.
As a result, the trajectory in case C
becomes similar to the one in case B,
which is ``compressed'' in the $z$-direction.

\begin{center}
------------ \\
Figure 1 \\
------------ \\
\end{center}

\begin{flushleft}
{\it 3.2. Accretion Radius}
\end{flushleft}

We  calculate the accretion radius by the almost same procedure
as Hoyle and Lyttleton (1939).
Let us suppose that two particles 
intersecting downstream on the accretion axis
collide and lose their transverse momenta.
If the resultant $z$-velocity is less than the escape velocity
at that point, these particles are assumed to be
captured and accreted on the gravitating body.
In case C, however, more sophisticated treatment is necessary.
During the coalesced particle moves on the accretion axis
after collision,
it is suffered from radiation drag and loses its $z$-momentum.
Thus, in order to determine the accretion radius,
we follows the particle motion on the accretion axis
whether it escapes or turns to infall toward the center
in case C.

Figure 2 shows the accretion radius $R_{\rm acc}$
normalized by the Hoyle-Lyttleton accretion radius $R_{\rm HL}$
($=2GM/v_{\infty}^2$)
as a function of the normalized luminosity $\Gamma$ ($=L/L_{\rm E}$)
in the case of $v_{\infty}=0.1c$.
As can be seen from figure 2, 
for case A (short-dashed line)
the normalized accretion radius $R_{\rm acc}/R_{\rm HL}$ 
is always unity,
since this is nothing but the original Hoyle-Lyttleton accretion
($R_{\rm acc}=R_{\rm HL}$).
In case B (chain-dotted line)
$R_{\rm acc}/R_{\rm HL}$ is a liner function of $\Gamma$
and it vanishes at $\Gamma=1$
[$R_{\rm acc}=R_{\rm HL}^{\rm R}=(1-\Gamma)R_{\rm HL}$].
Case C (solid curve) shows our present result.
The calculated $R_{\rm acc}/R_{\rm HL}$ in case C lies 
between those of cases A and B.
That is,
the effect of radiation drag increases
the accretion radius for case B without radiation drag.
The solid curve is the case
where the particle motion on the accretion axis
is followed, as stated above,
while the dashed one is the case
where the condition of accretion is determined
by the escape condition like cases A and B.
As is expected,
the effect of radiation drag on coalesed particles
slightly increases the accretion radius.

Figure 3 shows the accretion radius
for several cases of incident velocities $v_{\infty}$.
As can be seen in figure 3,
the effect of radiation drag becomes 
more important for a higher incident velocity
as large as $0.3c$.
The effect of radiation drag almost cancels 
the outward radiation pressure,
if $\Gamma$ is not so large.


\begin{center}
------------ \\
Figures 2-3 \\
------------ \\
\end{center}

We here derive the analytical approximate formula 
expressing the accretion radius for the case of small $v$.
If we restrict ourselves in the $R$-direction
and use $v$ instead of $v_R$,
equation (\ref{vector}) becomes
\begin{equation}
   \frac{dv}{dt} = -\frac{GM}{R^2} 
                   +\frac{\sigma_{\rm T}}{mc}
                    \left( F^R - 2Ev \right)
                 = -\frac{GM(1-\Gamma)}{R^2}
                   -\frac{GM}{R^2}2\Gamma\frac{v}{c}.
\end{equation}
The energy integration of this equation roughly gives
the following expression:
\begin{equation}
   \frac{1}{2}v_{\infty}^2 \sim \frac{GM(1-\Gamma)}{R_{\rm acc}}
              +\frac{GM}{R_{\rm acc}}2\Gamma\frac{v_{\infty}}{c}.
\end{equation}
Hence, the accretion radius under radiation drag
is roughly expressed as
\begin{equation}
   \frac{R_{\rm acc}}{R_{\rm HL}} \sim
      1 - \Gamma \left( 1 - \frac{2v_{\infty}}{c} \right).
\label{anaR}
\end{equation}
This formula (\ref{anaR}) well reproduces 
the results shown in figures 2-3 in the limit of small $v$, and
clearly shows the effects of radiation pressure and drag;
i.e., $\Gamma$ reduces $R_{\rm acc}$
while $v_\infty$ increases $R_{\rm acc}$.

\begin{flushleft}
{\bf 4. Concluding Remarks}
\end{flushleft}

We conclude that 
the accretion radius becomes larger,
if we take into account of radiation drag, 
compared with the case in which only the effect of radiation pressure
is taken into account.
The present effect becomes prominent when
the incident velocity of gas is comparable to the speed of light,
and the luminosity of the central object is comparable to
the Eddington luminosity.

In the present paper we bear in mind
the Compton drag in highly ionized media
(normal plasmas as well as electron-positron pair plasmas).
The result, however, is applicable
to more general cases including dust,
as long as the gray approximation is valid.
That is, the effect of radiation drag becomes important
if the specific cross-section of particles (``opacity'')
is sufficiently large.

We assume that the media is transparent.
In the hydrodynamical accretion, however,
the optical depth of the media should be considered.
For roughly uniform media, the optical depth exceeds unity
beyond some critical radius.
Hence, radiation drag as well as radiation pressure
effectively operate inside some radius.

In this paper
we consider the mildly relativistic case,
where the equations are of the first order of $v/c$.
In the case where the incident velocity is much close to $c$,
higher order terms should be included and
we must treat the problem 
within the framework of full special relativity.
In addition if the velocity is of the order of $c$,
the accretion radius becomes of the order of $r_{\rm g}$,
or the central source size for compact objects.
In such a situation
the effect of the finite source size
cannot be neglected.
In some cases
the general relativistic effect
becomes non-negligible.
These are left as a future work.


\vspace{0.5cm}
\noindent
{\bf Reference}\re
Bondi H. 1952, MNRAS 112, 195 \re
Bondi H., Hoyle F. 1944, MNRAS 104, 273 \re
Carroll D.L. 1990, ApJ 348, 588 \re
Davis R.E., Pringle J.E. 1980, MNRAS 191, 599 \re
Fortner B., Lamb F.K., Miller G.S. 1989, Nature 342, 775 \re
Fryxell B.A., Taam R.E., McMillan S.L.W. 1987, ApJ 315, 536 \re
Fukue J., Kato S., Matsumoto R. 1985, PASJ 37, 383 \re
Fukue J., Umemura M. 1995, PASJ 47, 429 \re
Guess A.W. 1962, ApJ 135, 855 \re
Ho C., Taam R.E., Fryxell B.A., Matsuda T., Koide H., Shima E.
   1989, MNRAS 238, 1447 \re
Hoyle F., Lyttleton R.A. 1939, Proc. Camb. Phil. Soc. 35, 405 \re
Hsieh S.-H., Spiegel E.A. 1976, ApJ 207, 244 \re
Hunt R. 1971, MNRAS 154, 141 \re
Kato S., Inagaki S., Fukue J., Mineshige S. (eds) 1996, 
   Basic Physics of Accretion Disks
   (Gordon Breach, New York) \re
Kato S., Fukue J., Mineshige S. 1998,
   Black-Hole Accretion Disks (Kyoto University Press, Kyoto) \re
Koide H., Matsuda T., Shima E. 1991, MNRAS 252, 473 \re
Meyer F., Meyer-Hofmeister E. 1994, A\&Ap 288, 175 \re
Poynting J.H. 1903, Phil Trans R Soc London, Ser A 202, 525 \re
Robertson H.P. 1937, MNRAS 97, 423 \re
Shima E., Matsuda T., Takeda H., Sawada K. 1985, MNRAS 217, 367 \re
Taam R.E., Fu A., Fryxell B.A. 1991, ApJ 371, 696 \re
Tajima Y., Fukue J. 1996, PASJ 48, 529 \re
Treves A., Maraschi L., Abramowicz M. 1988, PASP 100, 427 \re
Watanabe Y., Fukue J. 1996, PASJ 48, 849 \re

\newpage

\begin{center}
Figure Captions
\end{center}

Fig. 1.
Examples of particle trajectories
in the case of $v_{\infty}=0.1c$
for several values of $\Gamma$.
The values of $\Gamma$ are
(a) 0, (b) 0.5, (c) 1.0, and (d) 1.2.
The abscissae are $r$,
whereas the ordinates are $z$,
both in units of $r_{\rm g}$.
Particles, coming from upward initially,
are captured or scattered
due to the effects of 
gravity only (case A; short-dashed curves), 
gravity and radiation pressure (case B; chain-dotted curves), 
and gravity, radiation pressure, and radiation drag
(case C; solid curves).

Fig. 2.
Accretion radius normalized by the Hoyle-Lyttleton one,
$R_{\rm acc}/R_{\rm HL}$,
as a function of the normalized luminosity $\Gamma$.
The short-dashed horizontal line indicates case A,
where only the gravity is included.
The chain-dotted line, which linearly decreases with $\Gamma$,
expresses case B,
where the radiation pressure as well as gravity are included.
The thick solid curve shows  case C,
the present result under the effect of radiation drag.
The dashed curve also shows case C,
where the condition of accretion is determined
by the escape condition like cases A and B.

Fig.3.
Accretion radius for several values of
incident velocity $v_{\infty}$.

\newpage
\vspace*{30mm}
\leavevmode
\epsfxsize=60mm
\epsfbox{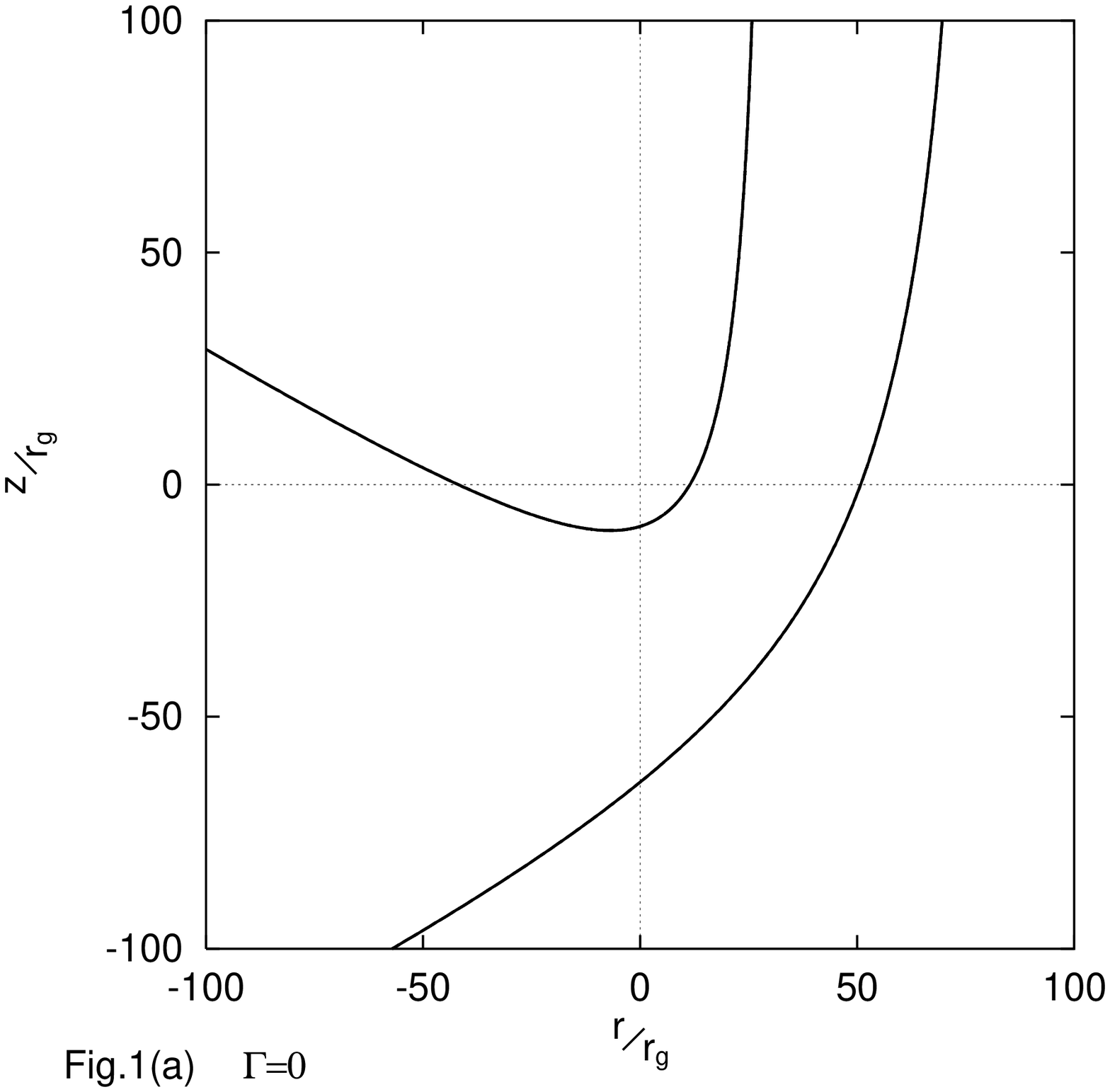}
\hspace{1cm}
\leavevmode
\epsfxsize=60mm
\epsfbox{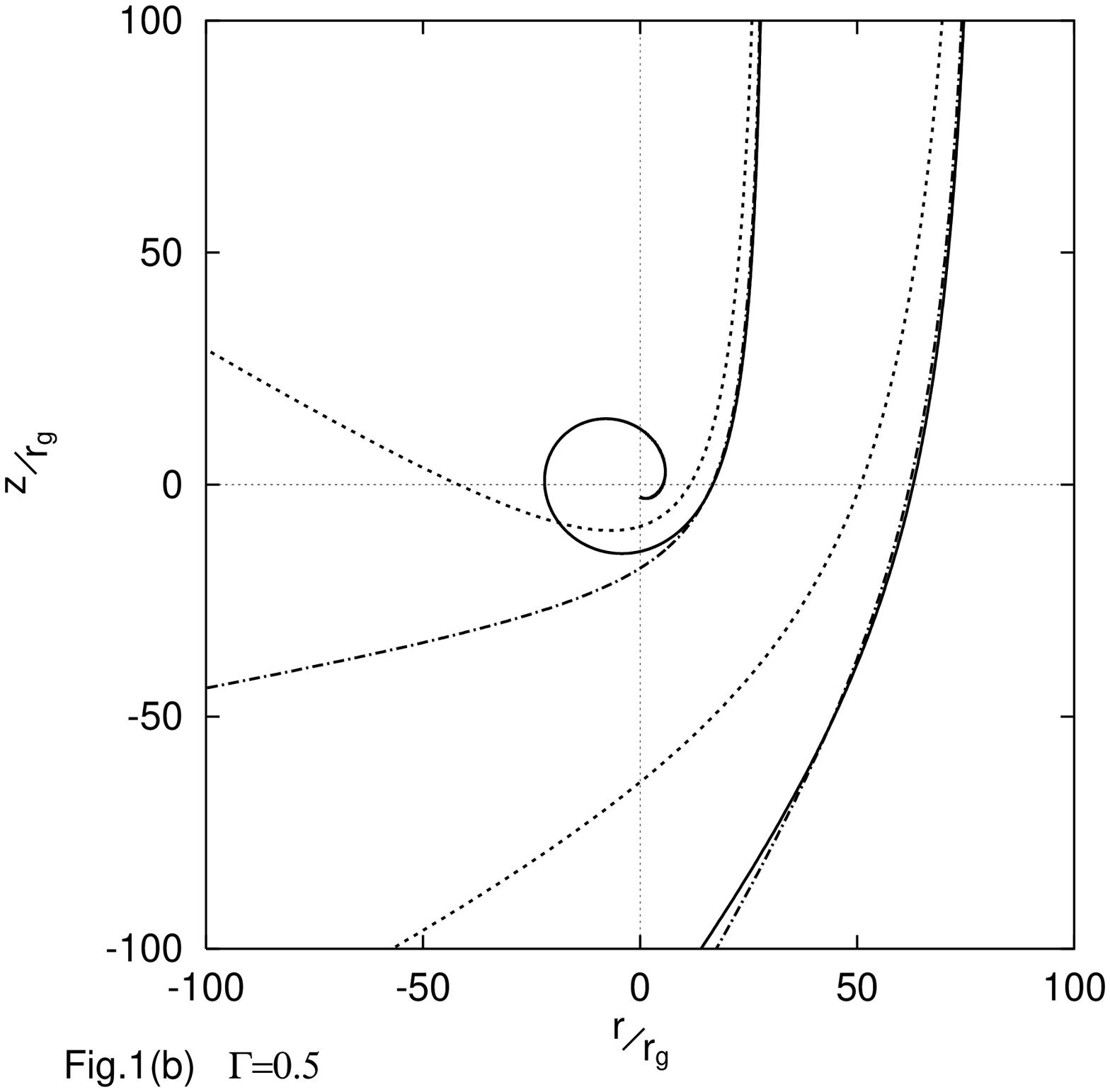}\\
\hspace*{5mm}
\vspace*{20mm}
\leavevmode
\epsfxsize=60mm
\epsfbox{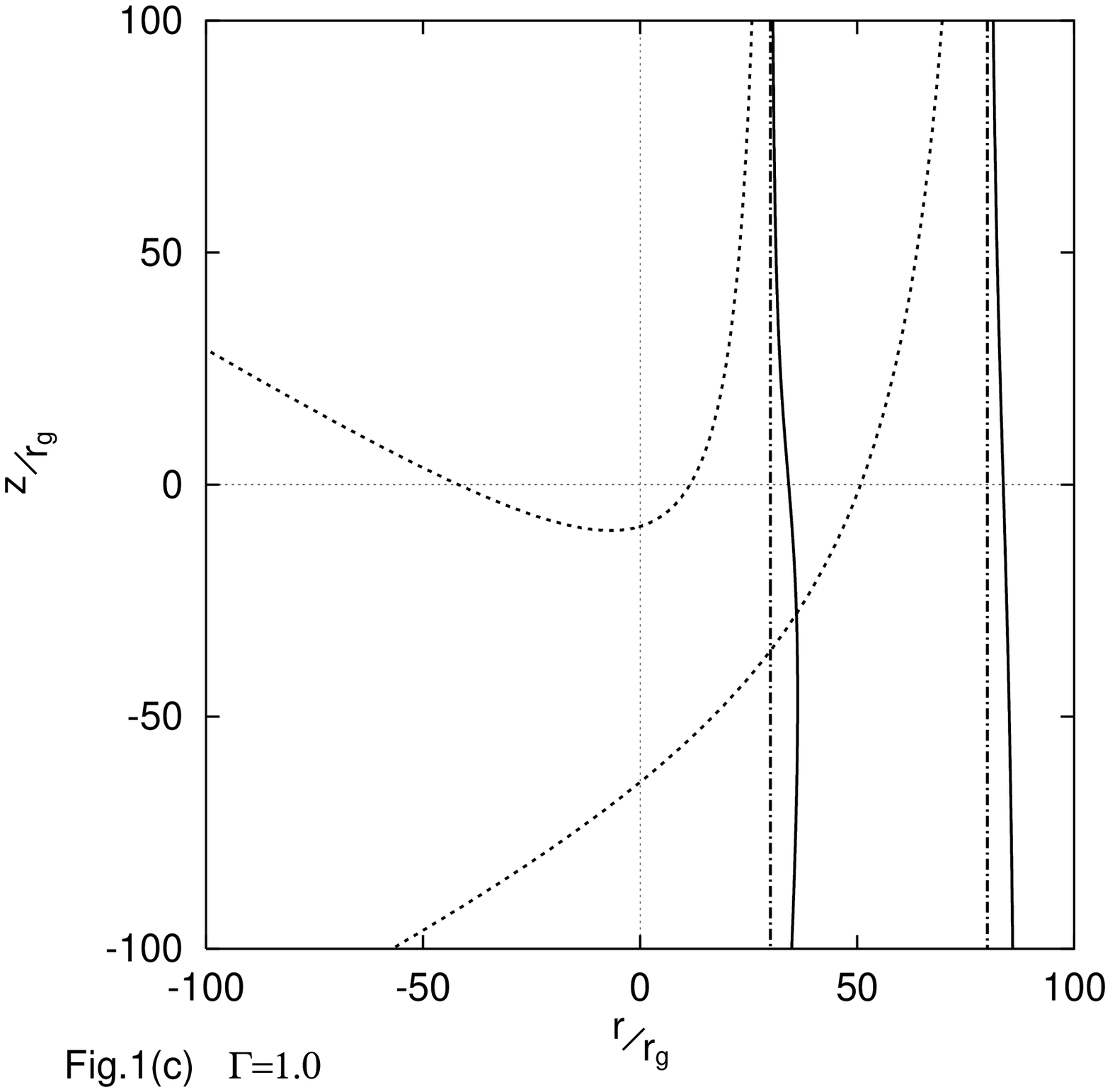}
\hspace{1cm}
\leavevmode
\epsfxsize=60mm
\epsfbox{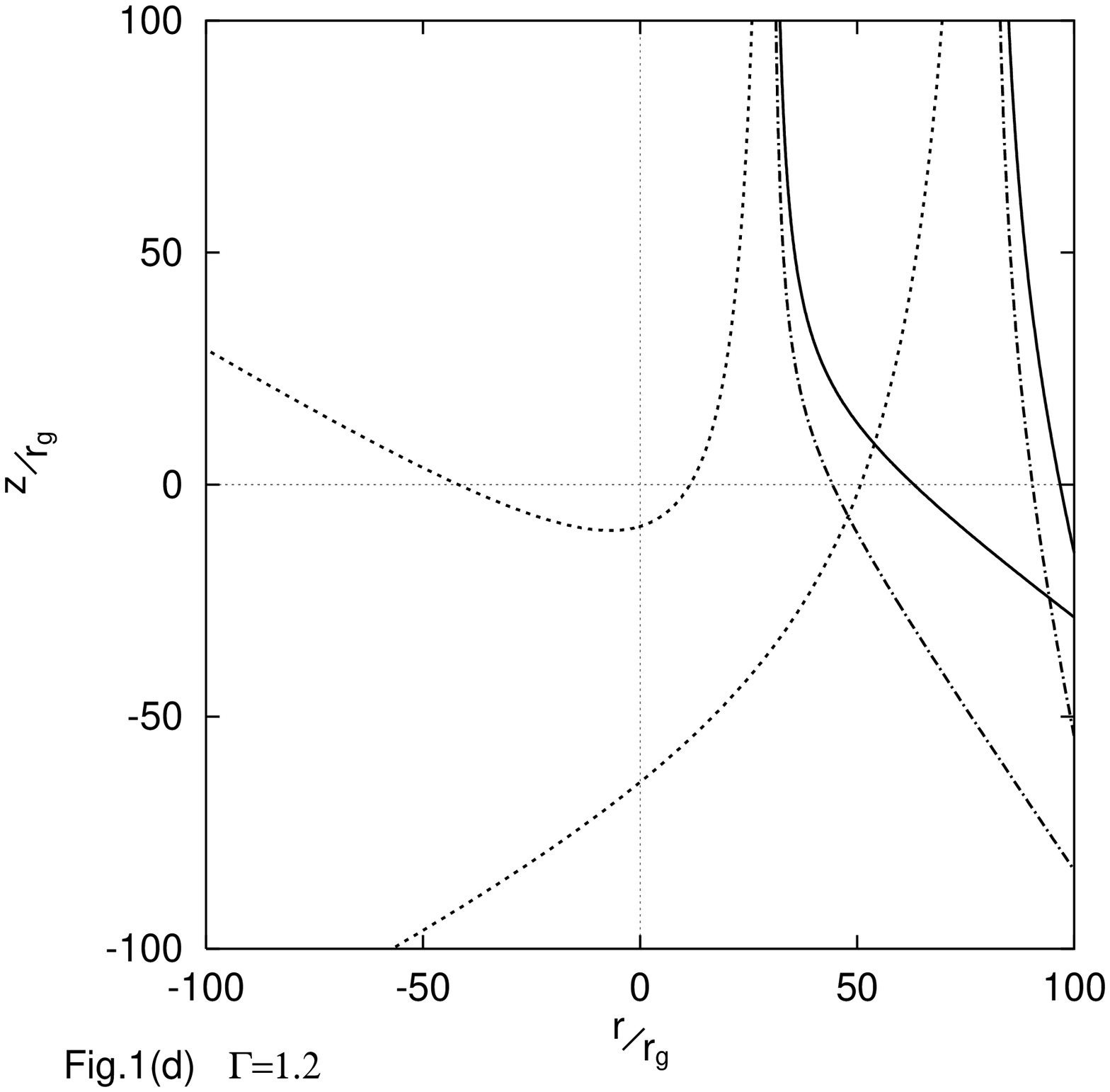}

\newpage
\begin{center}
\leavevmode
\epsfxsize=130mm
\epsfbox{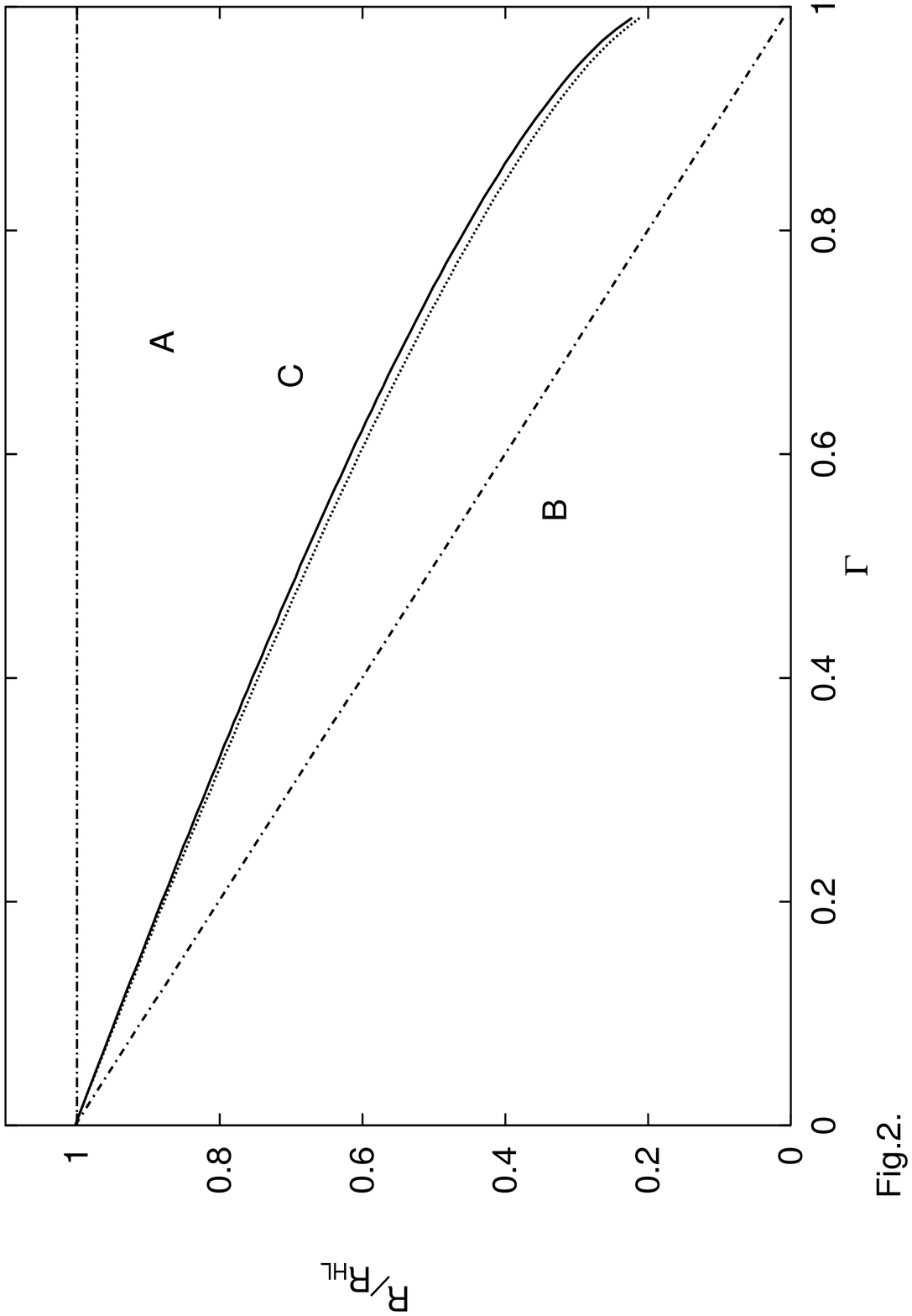}
\end{center}
\newpage
\begin{center}
\leavevmode
\epsfxsize=130mm
\epsfbox{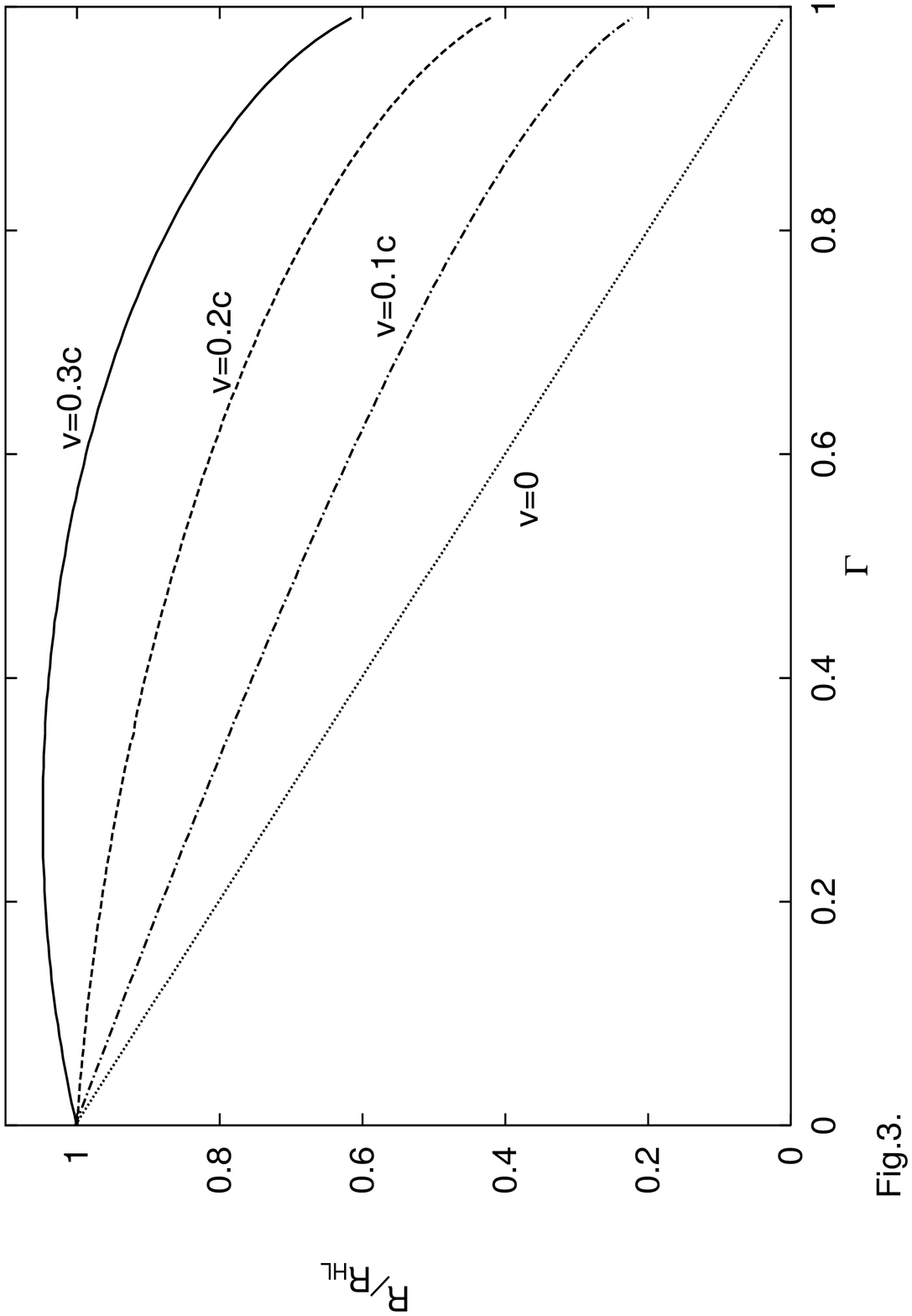}
\end{center}
\end {document}